\documentclass[10pt,conference]{IEEEtran}

\pdfoutput=1
\usepackage[caption=false]{subfig}
\usepackage[singlelinecheck=false]{caption} 

\usepackage{hyperref}
\usepackage{graphicx}
\usepackage{comment}
\usepackage{amsmath}
\usepackage{amssymb}
\usepackage{amsfonts}
\usepackage{bbm}
\usepackage{braket}
\usepackage{booktabs}
\usepackage{verbatim}
\usepackage{tabularx}
\usepackage{cite}
\usepackage{ulem}
\usepackage{color, colortbl}
\usepackage{multirow}
\definecolor{LightCyan}{rgb}{1,0.5,0.5}
\hypersetup{
  colorlinks   = true, 
  urlcolor     = blue, 
  linkcolor    = blue, 
  citecolor   = blue 
}

\usepackage{color}
\usepackage{tikz}

\newcommand{\be}{\begin{equation}}
\newcommand{\ee}{\end{equation}}

\newcommand{\gb}{{\vec{\gamma}, \vec{\beta}}}
\newcommand{\Z}{\hat{\sigma^z}}

\newcommand{\citefuture}[1]{}

\begin{document}

	\title{
	{Performance Evaluation and Acceleration of the QTensor Quantum Circuit Simulator on GPUs}
	}

\author{\IEEEauthorblockN{Danylo Lykov\IEEEauthorrefmark{1}\IEEEauthorrefmark{5}, Angela Chen\IEEEauthorrefmark{2}\IEEEauthorrefmark{5}, Huaxuan Chen\IEEEauthorrefmark{3}\IEEEauthorrefmark{5}, Kristopher Keipert\IEEEauthorrefmark{4}, Zheng Zhang\IEEEauthorrefmark{2}, Tom Gibbs\IEEEauthorrefmark{4}, and Yuri Alexeev\IEEEauthorrefmark{1}}
\IEEEauthorblockA{\IEEEauthorrefmark{1}Computational Science Division, Argonne National Laboratory, Lemont, IL 60439}
\IEEEauthorblockA{\IEEEauthorrefmark{2}Department of Electrical and Computer Engineering, University of California, Santa Barbara, CA 93106}
\IEEEauthorblockA{\IEEEauthorrefmark{3}Northwestern University, Evanston, IL 60208}
\IEEEauthorblockA{\IEEEauthorrefmark{4}NVIDIA Corporation, Santa Clara, CA 95050
}
\IEEEauthorblockA{\IEEEauthorrefmark{5} These authors contributed equally to the paper}
}

	\date{Started 6/3/21}

	\maketitle

\begin{abstract}
This work studies the porting and optimization of the tensor network simulator QTensor on GPUs, with the ultimate goal of simulating quantum circuits efficiently at scale on large GPU supercomputers. We implement NumPy, PyTorch, and CuPy backends and benchmark the codes to find the optimal allocation of tensor simulations to either a CPU or a GPU. We also present a dynamic mixed backend to achieve optimal performance. To demonstrate the performance, we simulate QAOA circuits for computing the MaxCut energy expectation. Our method achieves $176\times$ speedup on a GPU over the NumPy baseline on a CPU for the benchmarked QAOA circuits to solve MaxCut problem on a 3-regular graph of size 30 with depth $p=4$.
\end{abstract}

\section{Introduction}

Quantum information science (QIS) has a great potential to speed up certain computing problems such  as combinatorial optimization and quantum simulations \cite{alexeev2019quantumworkshop}. The development of fast and resource-efficient quantum simulators to classically simulate quantum circuits is the key to the advancement of the QIS field. For example, simulators allow researchers to evaluate the complexity of new quantum algorithms and to develop and validate the design of new quantum circuits. Another important application is to validate quantum supremacy and advantage claims.

One  can simulate quantum circuits on classical computers  in many ways. The major types of simulation approaches are full amplitude-vector evolution \cite{de2007massively, smelyanskiy2016qhipster, haner20170, wu2019full}, the Feynman paths approach \cite{bernstein1997quantum}, linear algebra open system simulation \cite{quac}, and tensor network contractions \cite{markov2008simulating, pednault2017breaking, boixo2017simulation}. These techniques have advantages and disadvantages. Some are better suited for small numbers of qubits and high-depth quantum circuits, while others are better for circuits with a large number of qubits but small depth. Some are also tailored toward the accuracy of simulation of noise in quantum computers.

For shallow quantum circuits the state-of-the-art technique to simulate quantum circuits is currently arguably the tensor network contraction method because of the memory efficiency for the method relative to state vector methods that scale by $2^N$, where $N$ is the number of qubits. This effectively limits the state vector methods to quantum circuits with less than 50 qubits. The challenge with the tensor network methods is determining the optimal contraction order, which is known to be an NP-complete problem~\cite{markov2008simulating}. We choose to focus on the simulation of the Quantum Approximate Optimization Algorithm (QAOA)~\cite{farhi2016quantum} given its importance to machine learning and its suitability for the current state of the art with noisy intermediate state quantum computers that generally work with circuits of short depth.


In this work we ported and optimized the tensor network quantum simulator QTensor to run efficiently on GPUs, with the eventual goal to simulate large quantum circuits on the modern and upcoming supercomputers. In particular, we benchmarked QTensor on a NVIDIA DGX-2 server with a V100 accelerator using the CUDA version 11.0. The performance is shown for the full expectation value simulation of the QAOA MaxCut problem on a 3-regular graph of size 30 with depth $p=4$.

\section{Methodology}

\subsection{QAOA Overview}

The Quantum Approximate Optimization Algorithm is a variational quantum algorithm that combines a parameterized ansatz state preparation with a classical outer-loop algorithm that optimizes the ansatz parameters.
QAOA is used for approximate solution of binary optimization problems~\cite{farhi2014quantum}.
A solution to the optimization problem is obtained by measuring the ansatz state on a quantum device.
The quality of the QAOA solution depends on the depth of the quantum circuit that generated the ansatz and the quality of parameters for the ansatz state.

A binary combinatorial optimization problem is defined on a space of binary strings of length $N$ and has $m$ clauses.
Each clause is a constraint satisfied by some assignment of the bit string.
QAOA maps the combinatorial optimization problem onto a $2^N$-dimensional Hilbert space with computational basis vectors $\ket{z}$ and encodes $C(z)$ as a quantum operator $\hat C$ diagonal in the computational basis.
One of the most widely used benchmark combinatorial optimization problems is MaxCut, which is defined on an undirected
unweighted graph.
The goal of the MaxCut problem is to find a partition of the graph's vertices into two complementary sets 
such that the number of edges between the sets is maximized.
It has been shown in \cite{farhi2014quantum} that on a 3-regular graph, QAOA with $p=1$ produces a solution with an approximation ratio of at least 0.6924.

A graph $G = (V, E)$ of $N = |V|$ vertices and $m=|E|$ edges can be encoded into a MaxCut cost operator
over $N$ qubits by using $m$ two-qubit gates.
\begin{equation} \label{eq:maxcut_cost}
    \hat C = \frac{1}{2} \sum_{(ij)\in E} 1-\Z_i\Z_j
\end{equation}

The QAOA ansatz state $\ket \gb$ is prepared by applying $p$ layers of evolution unitaries that correspond to 
the cost operator $\hat C$ and a mixing operator $\hat B = \sum_{i\in V} \hat \sigma^x_i$.
The initial state is the equally weighted superposition state and maximal eigenstate of $\hat B$.
\begin{equation}
    \ket{\gb}_p = \prod_{k=1}^p e^{-i\beta_k\hat B} e^{-i\gamma_k\hat C} \ket{+}
    \label{eq:ansatz_state}
\end{equation}
The parameterized quantum circuit (\ref{eq:ansatz_state}) is called the \emph{QAOA ansatz}. We refer to the number of alternating operator pairs~$p$ as the \emph{QAOA depth}. 

The solution to the combinatorial optimization problem is obtained by measuring the QAOA ansatz.
The expected quality of this solution is an expectation value of the cost operator in this state.

\[\langle C\rangle_p = \bra{\gb}_pC\ket{\gb}_p 
\]
The expectation value can be minimized with respect to parameters $\gb$.
The optimization of $\gb$ can be performed by using classical computation or by varying the parameters and sampling many bitstrings from a quantum computer
to estimate the expectation value.
Acceleration of the optimal parameters search for a given QAOA depth $p$ is the focus of many approaches aimed at demonstrating the quantum advantage. Examples include such methods as warm- and multistart  optimization~\cite{egger2020warmstarting,shaydulin2019multistart}, problem decomposition~\cite{shaydulin2019hybrid},  instance structure analysis~\cite{shaydulin2020classical}, and parameter learning~\cite{khairy2020learning}.

In this paper we focus on application of a classical quantum circuit simulator QTensor to the problem
of finding the expectation value $\langle C \rangle_p$.

\subsection{Tensor Network Contractions}

Calculation of an expectation value of some observable in a given state generated by some
quantum circuit can be done efficiently by using a
tensor network approach.
In contrast to state vector simulators, which store the full state vector of size $2^N$, QTensor maps a quantum circuit to a tensor network. Each quantum gate of the circuit is converted to a tensor.
An expectation value $\braket{\phi| \hat C | \phi} = \braket{\psi | \hat U^\dagger \hat C \hat U| \psi}$ is then simulated by contracting the corresponding tensor network. For more details on how a quantum circuit is converted to a tensor network, see \cite{Schutski_2020, lykov2020tensor}.

A tensor network is a collection of tensors, which in turn have a collection of indices, where tensors share some indices with each other.
To contract a tensor network, we create an ordered list of tensor buckets.
Each bucket (a collection of tensors) corresponds to a tensor index, which is called \emph{bucket index}.
Buckets are then contracted one by one.
The contraction of a bucket is performed by summing over the bucket index, and the resulting
tensor is then appended to the appropriate bucket.
The number of unique indices in aggregate indices of all bucket tensors is called a \emph{bucket width}.
The memory and computational resources of a bucket contraction scale exponentially with the associated bucket width.
For more information on tensor network contraction, see  \cite{lykov2021importance, lykov2021large, shutski2019adaptive}.
If some observable $\hat \Sigma$ acts on a small subset of qubits, most of the gates in the quantum circuit $\hat U$ cancel out when evaluating the expectation value.
The cost QAOA operator $\hat C$ is a sum of $m$ such terms, each of which could be viewed as a separate observable.
Each term generates a \emph{lightcone}---a subset of the problem that
generates a tensor network representing the contribution to the cost expectation value.

The expectation value of the cost for the graph $G$ and MaxCut QAOA depth $p$ is then
\begin{align*}
\braket{C}_p(\gb) &= \braket{\gb| \hat C | \gb} 
\\
&= \braket{\gb| \sum_{jk\in E}\frac{1}{2}(1-\Z_j\Z_k)|\gb}
\\
&= \frac{|E|}{2} - \frac{1}{2}\sum_{jk\in E}\braket{\gb|\Z_j\Z_k|\gb}
\\
&\equiv \frac{|E|}{2} - \frac{1}{2}\sum_{jk\in E}e_{jk}(\gb),
\end{align*}
where $e_{jk}$ is an individual edge contribution to the total cost function.
Note that the observable in the definition of $e_{jk}$ is local to only two qubits; therefore most of the gates in the circuit that generates the state $\ket \gb$ cancel out. The circuit after the cancellation is equivalent to calculating $\Z_j\Z_k$ on a subgraph $S$ of the original graph $G$. 
These subgraphs can be obtained by taking only the edges that are incident from vertices at a distance $p-1$ from the vertices $j$ and $k$.
The full calculation of $E_G(\gb)$ requires evaluation of $|E|$ tensor networks, each representing the value $e_{jk}(\gb)$ for $jk \in E$.

\subsection{Merged Indices Contraction}
\label{sec:merged}

Since the contraction in the bucket elimination algorithm is executed one index at a time, the ratio of computational operations to memory read/write operations is small.
This ratio is also called the operational intensity or arithmetic intensity. Having small arithmetic intensity hurts the performance in terms of FLOPs: for each floating-point operation calculated there are relatively many I/O operations, which are usually slower.
For example, to calculate one element of the resulting matrix in a matrix multiplication problem, one needs to read $2N$ elements and perform $4N$ operations. The size of the resulting matrix is similar to the input matrices.
In contrast, when calculating an outer product of two vectors, the size of the resulting matrix is much larger than the combined size of the input vectors; each element requires two reads and only one floating-point operation.

To mitigate this limitation, we develop an approach for increasing the  arithmetic intensity, which we call merged indices. The essence of the approach is to combine several buckets and contract their corresponding indices at once, thus having smaller output size and larger arithmetic intensity. 
We have a group of circuit contraction backends that all use this approach.

For the merged backend group, we order the buckets first and then find the mergeable indices before performing the contraction. We list the set of indices of tensors in each bucket and then merge the buckets if the set of indices of one bucket is a subset of the other. We benchmark the sum of the total time needed for the merged indices contraction and compare it with the unmerged baseline results. We call this group the “merged” group and the baseline the “unmerged” group.

\subsection{CPU-GPU Hybrid Backend}
The initial tensor network contains only very small tensors of at most 16 elements (4 dimensions of size 2). We observe that the contraction sequence obtained by our ordering algorithm results in buckets of small width for first 80\% of contraction steps. Only after all small buckets are contracted, sequence we start to contract large buckets. The GPUs usually perform much better when processing large amount of data. We observe this behaviour in our benchmarks on Figure \ref{fig:meantime_bucket}.
We therefore implement a mix backend which uses both CPU and GPU. It combines a CPU backend and a GPU backend by dispatching the contraction procedure to appropriate backend.

The mix or the hybrid backend uses the bucket width, which is determined by the number of unique indices in a bucket, to allocate the correct device for such a bucket to be computed. The threshold between the CPU backend and the GPU backend is determined by a trial program. This program runs a small circuit, which is used for all backends for testing, separately on a GPU backend and a CPU backend. After the testing is complete, it iterates through all bucket widths and checks whether at this bucket width the GPU takes less time or not. If it finds the bucket width at which the GPU is faster, it will output that bucket width, and the user can use this width when creating the hybrid backend in the actual simulation. In the actual simulation, if the bucket width is smaller than the threshold, the hybrid backend will allocate this bucket to the CPU and will allocate it to the GPU if the width is greater.

Since we don't contract buckets of large width on CPU, the resulting tensors are rather small, on the order of 1,000s of bytes. The time for data transfer in this case is considered negligible and is not measured in our code. The large tensors start to appear from contractions that combine these small tensors after all the data is moved to GPU.

\subsection{Datasets for Synthetic Benchmarks}
\label{sec:methodology_synthetic}

Tensor network contraction is a complex procedure that involves many inhomogeneous operations. 
Since we are interested in achieving the maximum performance of the simulations, it is beneficial to compare the FLOPs performance to several more relevant benchmarking problems.
We select several problems for this task:
\begin{enumerate}
    \item Square matrix multiplication, the simplest benchmark problem which serves as an upper bound for our FLOP performance;
    \item Pairwise tensor contractions with a small number of large dimensions and fixed contraction structure;
    \item Pairwise tensor contractions with a large number of dimensions of size 2 and permuted indices;
    \item Bucket contraction of buckets that are produced by actual expectation value calculation;
    \item Full circuit contraction which takes into account buckets of large and small width.
\end{enumerate}
By gradually adding complexity levels to the benchmark problems and evaluating the performance on each level, we look for the largest reduction in FLOPs.
The corresponding level of complexity will be at the focus of our future efforts for optimisation of performance.
The results for these benchmarks are shown in Section \ref{sec:results_synthetic} and Figures \ref{fig:tncontract_cupy} and \ref{fig:tncontract_numpy}.

\subsubsection{Matrix Multiplication}
We perform the matrix multiplications for the square matrices of the same size and record the time for the operation for the CPU backend Numpy and the GPU backends PyTorch and CuPy. We use the built-in \texttt{random()} function of each backend to randomly generate two square matrices of equal size as our input, and we use the built-in \texttt{matmul()} function to produce the output matrix. The size of the input matrices ranges from $10 \times 10$ to $8192 \times 8192$, and the test is done repeatedly on four different data types: \texttt{float}, \texttt{double}, \texttt{complex64}, and \texttt{complex128}. For the multiplication of two $n \times n$ matrices, we define the number of complex operations to be $n^3$, and we calculate the number of FLOPs for complex numbers as $8\times\frac{\text{number\_of\_operations}}{\text{operation\_time}}$.

\subsubsection{Tensor Network Contraction}
We have two experiment groups in benchmarking the tensor contraction performance: tensor contractions with a fixed contraction expression and tensor contractions with many indices where each index has a small size. We call the former group “tncontract fixed” because we fix the contraction expression as “abcd,bcdf$\xrightarrow{}$acf,” and we call the latter one “tncontract random” because we randomly generate the contraction expression. In a general contraction expression, we sum over the indices not contained in the result indices. In this fixed contraction expression, we sum over the common index “b” and “d” and keep the rest in our result indices. We generate two square input tensors of shape $n \times n \times n \times n$ and output a tensor of shape $n \times n \times n$, where $n$ is a size ranging from 10 to 100. For the “tncontract random” group, we randomly generate the number of contracted indices and the number of indices in the results first and then fill in the shape array with size 2. For example, a contraction formula “dacb,ad$\xrightarrow{}$bcd” (index “a” is contracted) needs two input tensors: the first one with shape $2 \times 2 \times 2 \times 2$ and the second one with shape $2 \times 2$. We use the formula $2^{\text{number\_of\_different\_indices}}$ to calculate the number of operations, and we record the contraction time and compute the FLOPs value based on the formula used in matrix multiplication. Following the same procedure in matrix multiplication, we use the backends’ built-in functions to randomly generate the input tensors based on the required size and the four data types.

\subsubsection{Circuit Simulation}
For numerical evaluations, we benchmark the full expectation value simulation of the QAOA MaxCut problem for a 3-regular graph of size 30 and QAOA depth $p=4$. We have two properties for evaluating the circuit simulation performance: unmerged vs. merged backend and single vs. mixed backend.

\section{Results}
\begin{figure}
    \centering
    \includegraphics[width=\linewidth]{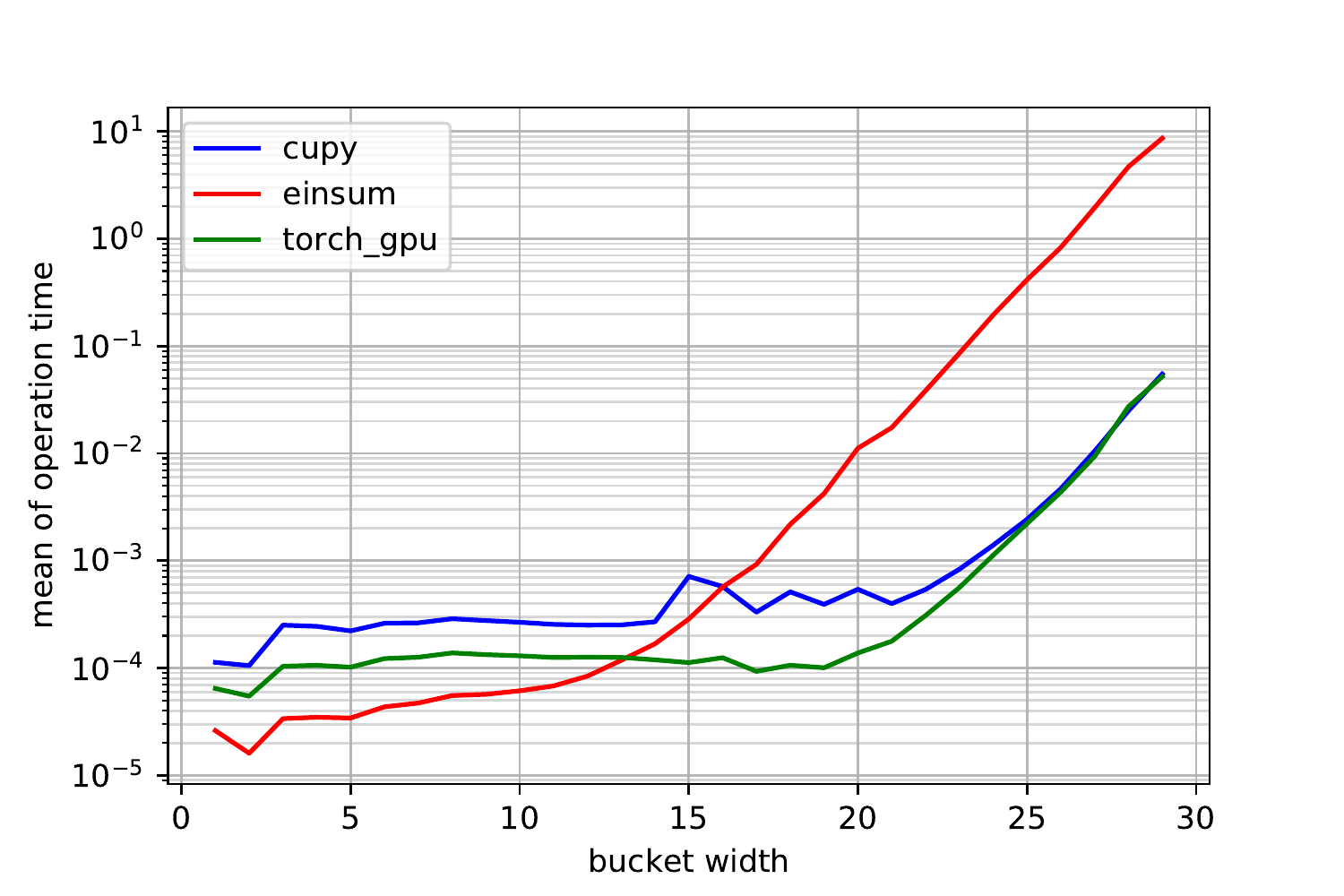}
    \caption{
    Breakdown of mean time to contract a single bucket by bucket width. The test is performed for expectation value as described in \ref{sec:single_backends}.
    CPU backends are faster for buckets of width~$\leq13-16$, and GPU faster
    are better for larger buckets. 
    This picture also demonstrates that every contraction operation spends some time on overhead which doesn't depend on bucket width, and actual calculation that scales exponentially with bucket width.
    }
    \label{fig:meantime_bucket}
\end{figure}

\begin{figure}
    \centering
    \includegraphics[width=\linewidth]{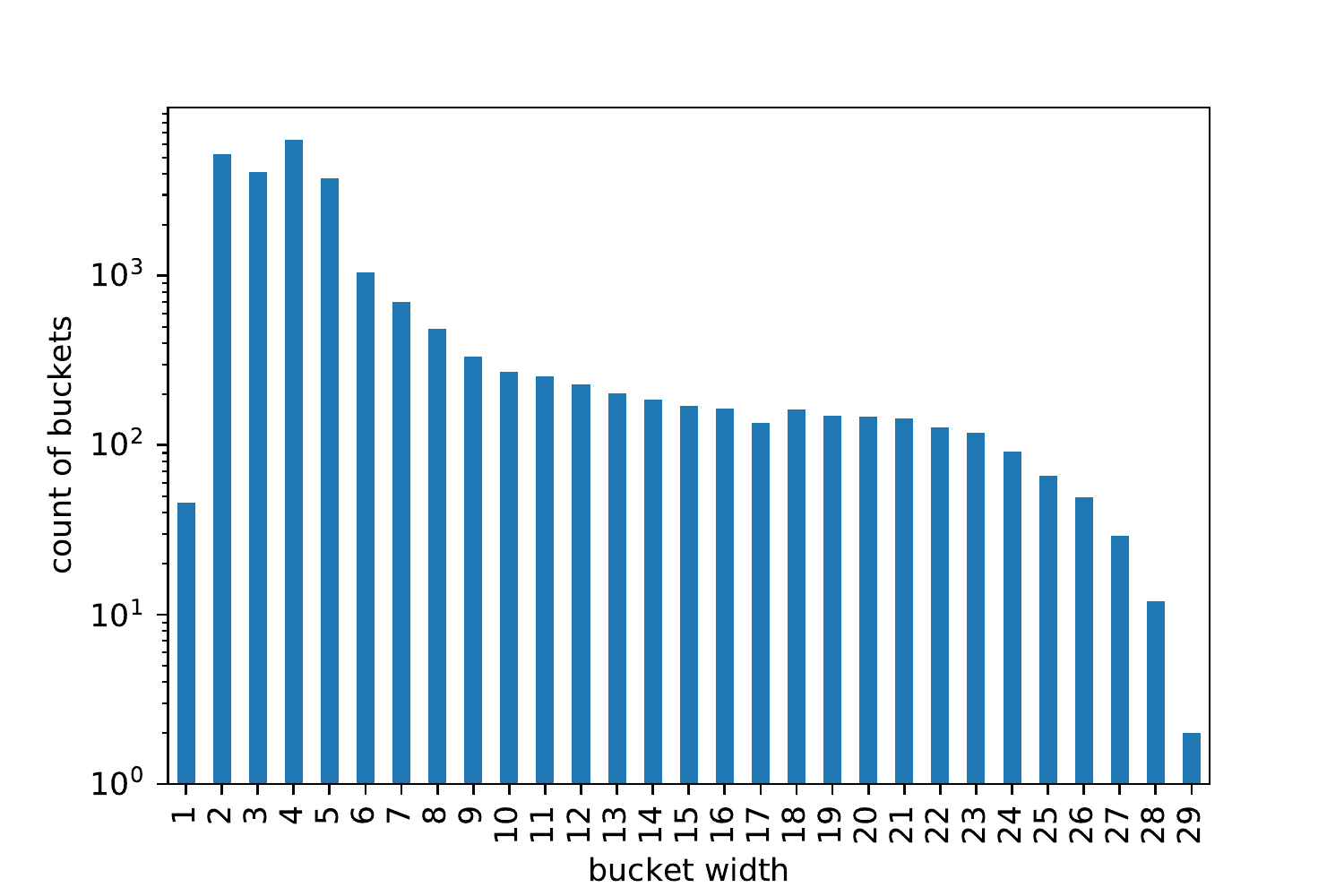}
    \caption{Distribution of bucket width in the contraction of QAOA full circuit simulation. The y-axis is log scale; 82\% of buckets have width $\leq 6$, which have relatively large overhead time.}
    \label{fig:count_bucket}
\end{figure}

\begin{figure}
    \centering
    \includegraphics[width=\linewidth]{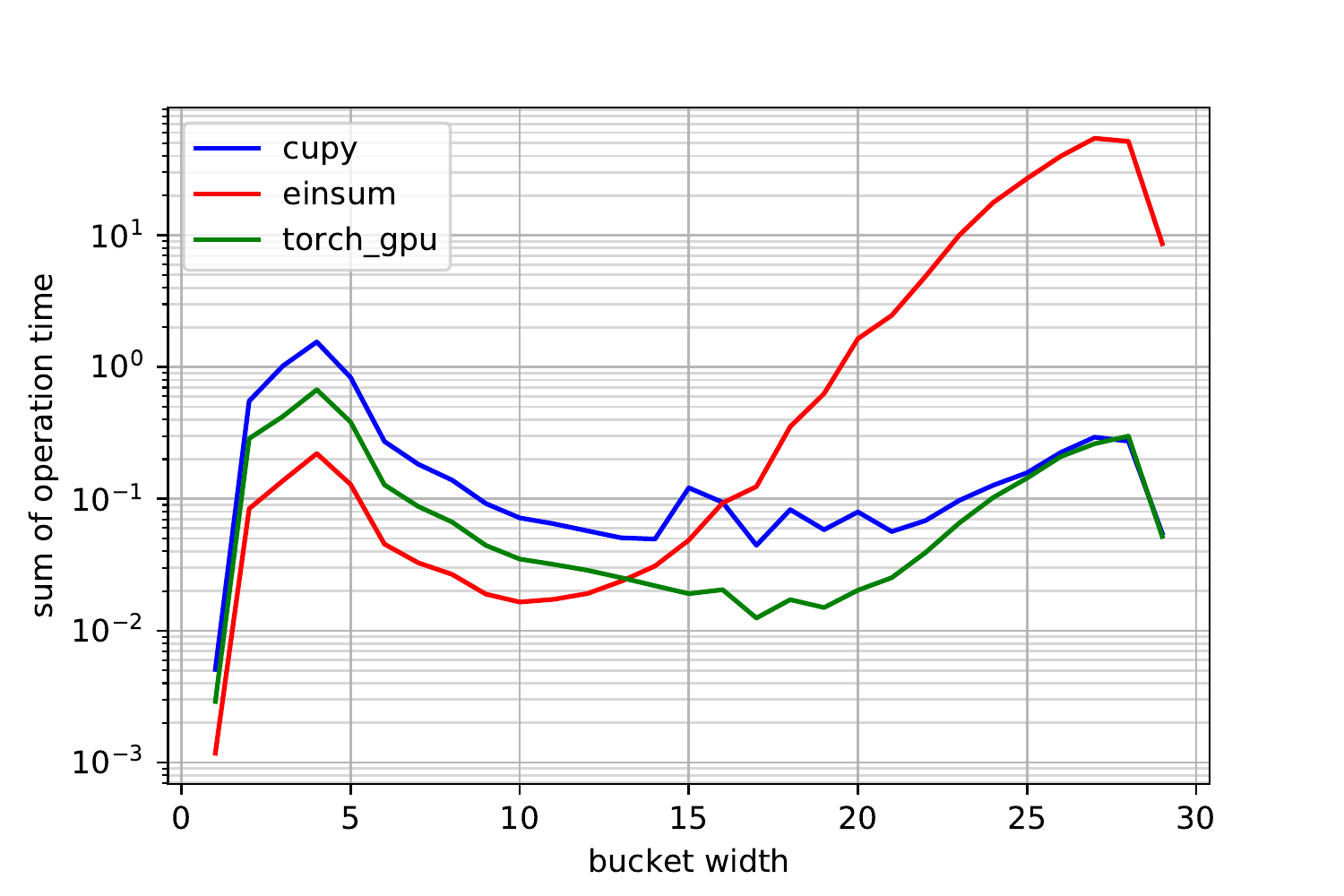}
    \caption{
    Breakdown of total time spent on bucket of each size in full QAOA expectation value simulation.
    The y-value on this plot is effectively one in Figure~\ref{fig:meantime_bucket} multiplied by one in Figure~\ref{fig:count_bucket}.
    This figure is very useful for analyzing the bottlenecks of the simulation. It shows that most of the time for CPU backend is spent on large buckets, but for GPU backends the large number of small buckets results in a slowdown.
    }
    \label{fig:time_bucket}
\end{figure}

The experiment is performed on an NVIDIA DGX-2 server (provided by NVIDIA corporation) with a V100 accelerator using the CUDA version 11.0. The baseline NumPy backend is executed only on a CPU and labeled ``einsum" in our experiment since we use \texttt{numpy.einsum()} for the tensor computation. 
We also benchmark the GPU library CuPy (on the GPU only) and PyTorch (on both the CPU and GPU).

\subsection{Single CPU-GPU Backends}
\label{sec:single_backends}



We benchmark the performance of the full expectation value simulation of the QAOA MaxCut problem on a 3-regular graph of size 30 with depth $p=4$, as shown in in Figures \ref{fig:meantime_bucket}, \ref{fig:count_bucket}, and \ref{fig:time_bucket}. This corresponds to contraction of 20 tensor networks, one network per each lightcone. Our GPU implementation of the simulator using PyTorch (labeled ``torch\_gpu") achieves 70.3$\times$ speedup over the CPU baseline and 1.92$\times$ speedup over CuPy.

Figure \ref{fig:meantime_bucket} shows the mean contraction time of various bucket widths in different backends. In comparison with "cupy" backend, the "einsum" backend spends less total time for bucket width less than 16, and the threshold value changes to around 13 when being compared to "torch\_gpu" backend. Both GPU backends have similar and better performance for larger bucket widths. However, this threshold value can fluctuate when comparing the same pair of CPU and GPU backends. This is likely due to the fact that the benchmarking platform are under different usage loads.

Figure \ref{fig:time_bucket} provides a breakdown of the contraction times of buckets by bucket width. This distribution is multimodal: A large portion of time is spent on buckets of width 4. For CPU backends the bulk of the simulation time is spent on contracting large buckets. Figure \ref{fig:count_bucket} shows the distribution of bucket widths, where 82\% of buckets have width less than 7. This signifies that simulation has an overhead from contracting a large number of small buckets.

This situation is particularly noticeable when looking at the total contraction time of different bucket widths. Figure \ref{fig:time_bucket} shows that the distribution of time vs bucket width has two modes: for large buckets that dominate the contraction time for CPU backends and for small buckets where most of the time is spent on I/O and other code overhead.

\subsection{Merged Backend Results}

\begin{figure}
    \centering
    \includegraphics[width=\linewidth]{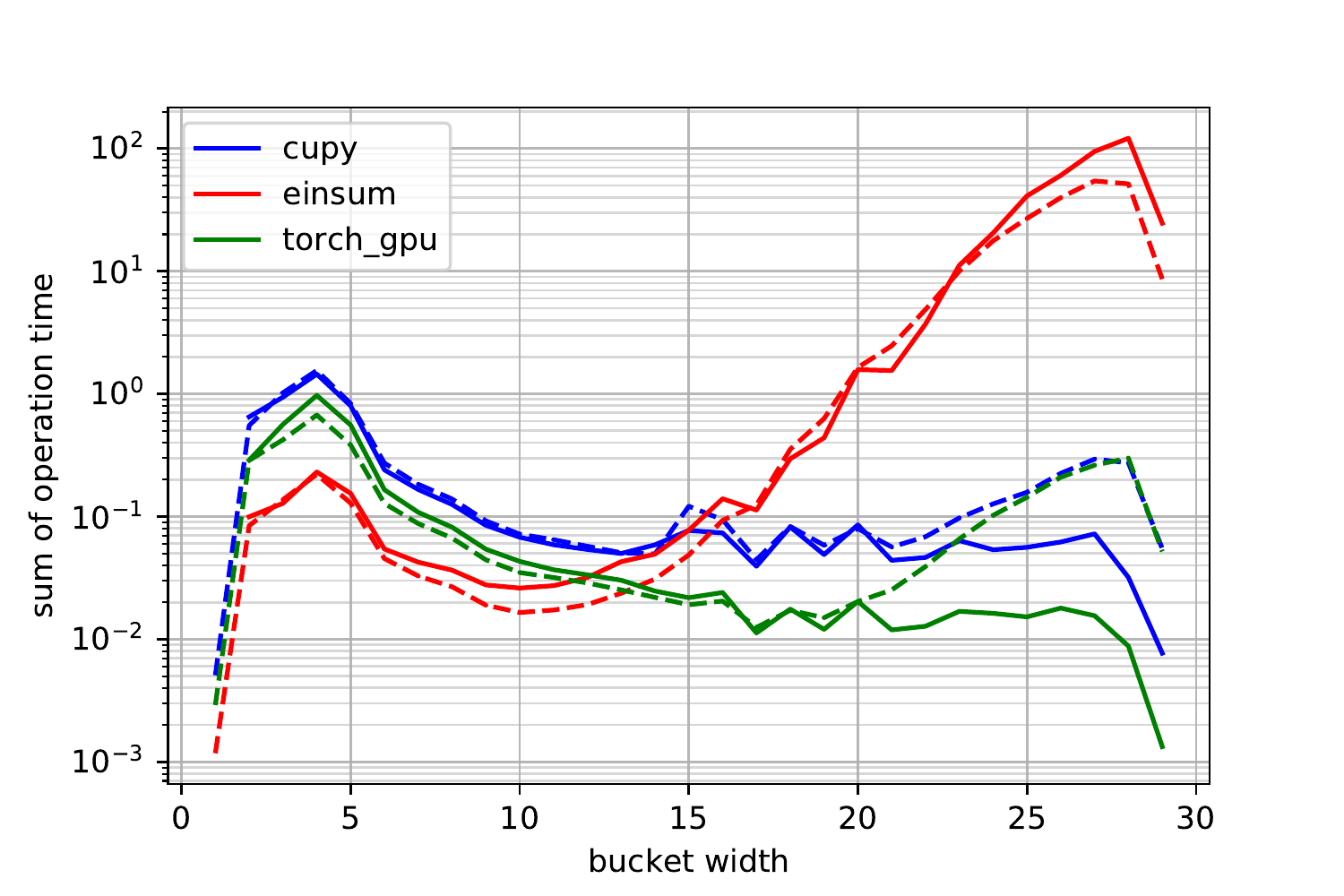}
    \caption{
    Breakdown of total contraction time by bucket width in full expectation value simulation of problem size 30. Lines with the same color use the same type of backends. The solid lines represent the merged version of backends, and the dashed lines denote the baseline backends.
    The merged GPU backends are better for buckets of width $\geq 20$.
    }
    \label{fig:merged}
\end{figure}

The “merged” groups merge the indices before performing contractions. In Fig. \ref{fig:merged}, the three unmerged (baseline) backends are denoted by dashed lines, while the merged backends are shown by solid lines. For the GPU backends CuPy and PyTorch, the merged group performs significantly better for buckets of width $\geq 20$. The CuPy merged backend always has a similar or better performance compared with the CuPy unmerged group and has much better performance for buckets of larger width. For buckets of width 28, the total operation time of the unmerged GPU backends is about 0.28 seconds, compared with 32 ms (8.75$\times$ speedup) for the CuPy merged group and 8.8 ms (31.82$\times$ speedup) for the PyTorch backend. But we do not observe much improvement for the merged CPU backend.

\subsection{Mix CPU-GPU Backend Results}

\begin{table}[]
    \centering
        \begin{tabular}{|l|l|l|l|}
        \hline
         \textbf{Backend Name}        & \textbf{Device}  & \textbf{Time (second)} & \textbf{Speedup}\\
         \hline
         Torch\_CPU                 & CPU   & 347   & 0.71$\times$\\
         \textbf{NumPy} (baseline)  & CPU   & {246} & 1.00$\times$\\
         CuPy                       & GPU   & 6.7   & 36.7$\times$\\
         Torch\_GPU                 & GPU   & 3.5   & 70.3$\times$\\
         Torch\_CPU + Torch\_GPU    & Mixed & 2.6   & 94.8$\times$\\
         NumPy + CuPy               & Mixed & 2.1   & \textbf{117}$\times$\\
         \hline
    \end{tabular}
    \vspace{.5em}
    \small
    \caption{\small
    Time for full QAOA expectation value simulation using backend that utilize GPUs or CPUs. The expectation value is MaxCut on a 3-regular graph of size 30 and QAOA depth $p=4$.
    \textbf{Speedup} shows the overall runtime improvement compared with the baseline CPU backend ``NumPy". ``Mixed" device means the backend uses both CPU and GPU devices.
    }
    \label{tab:time}
\end{table}


From Figure \ref{fig:meantime_bucket} one can see that GPU backends perform much better
for  buckets of large width, while CPU backends are better for smaller buckets.
We thus implemented a mixed backend approach, which
dynamically selects a device (CPU or GPU) on which the bucket should be contracted.
We select a threshold value of 15 for the bucket width; any bucket that has 
a width larger than 15 will be contracted on the GPU.
Figure \ref{fig:time_bucket} shows that for GPU backends small buckets occupy
approximately 90\% of the total simulation time.
The results for this approach are shown in Table \ref{tab:time} under backend names
``Torch\_CPU + Torch\_GPU" and ``NumPy + CuPy."
Using a CPU backend in combination with Torch\_GPU improves the performance by 1.2$\times$,
and for CuPy the improvement is 3$\times$.
These  results suggest that using a combination of NumPy + Torch\_GPU
has the potential to give the best results.

We have evaluated the GPU performance of tensor network contraction 
for the energy calculation of QAOA. The problem is 
largely inhomogeneous with a lot of small buckets and a few very large buckets.
Most of the improvement comes from using GPUs on large buckets, with up to 300$\times$ speed improvement.
On the other hand, the contraction of smaller tensors is faster on CPUs.
In general, if the maximum bucket width of a lightcone is less than $\sim 17$, the
improvement from using GPUs is  marginal.
In addition, large buckets require a lot of memory.
For example, a bucket of width 27 produces a tensor with 27 dimensions of size 2, and 
the memory requirement for \texttt{complex128} data type is 2 GB.
In practice, these calculations are feasible up to width $\sim29$.

\subsection{Mixed Merged Backend Results}

Since the performance of the NumPy-CuPy hybrid backend is the best among all implemented hybrid backends, cross-testing between
merged backends and hybrid backends focuses on the combination of the NumPy backend and CuPy backend. Because of the API constraint, the hybrid of a regular NumPy backend and a merged CuPy backend was not implemented.

\begin{table}[]
    \centering
        \begin{tabular}{|l|l|l|l|}
        \hline
         \textbf{Backend Name}        & \textbf{Device}  & \textbf{Time (seconds)} & \textbf{Speedup}\\
         \hline
         NumPy\_Merged              & CPU   & 383   & 0.64$\times$\\
         \textbf{NumPy} (baseline)  & CPU   & {246} & 1.00$\times$\\
         CuPy                       & GPU   & 6.7   & 36.7$\times$\\
         CuPy\_Merged               & GPU   & 5.6   & 43.9$\times$\\
         NumPy + CuPy               & Mixed & 2.1   & 117$\times$\\
         NumPy\_Merged + CuPy\_Merged              & Mixed & 1.4   & \textbf{176}$\times$\\
         \hline
    \end{tabular}
    \vspace{.5em}
    \small
    \caption{\small
    Time for full QAOA expectation value simulation using different Merged backends, as described in Section \ref{sec:merged}. The expectation value is MaxCut on a 3-regular graph of size 30 and QAOA depth $p=4$.
    \textbf{Speedup} shows the overall runtime improvement compared with the baseline CPU backend ``NumPy".
    }
    \label{tab:time for MM}
\end{table}

In Table \ref{tab:time for MM}, merging buckets provide a performance boost for the CuPy backend and Numpy + CuPy hybrid backend but not the NumPy backend. CuPy\_Merged is 20\% faster than CuPy, and NumPy\_Merged + CuPy\_Merged is 50\% faster than its regular counterpart. However, NumPy\_Merged has an significant slowdown compared  with the baseline NumPy, suggesting that combining the  regular NumPy backend with the merged CuPy backend can provide more speedup for the future.

\begin{figure}
    \centering
    \includegraphics[width=\linewidth]{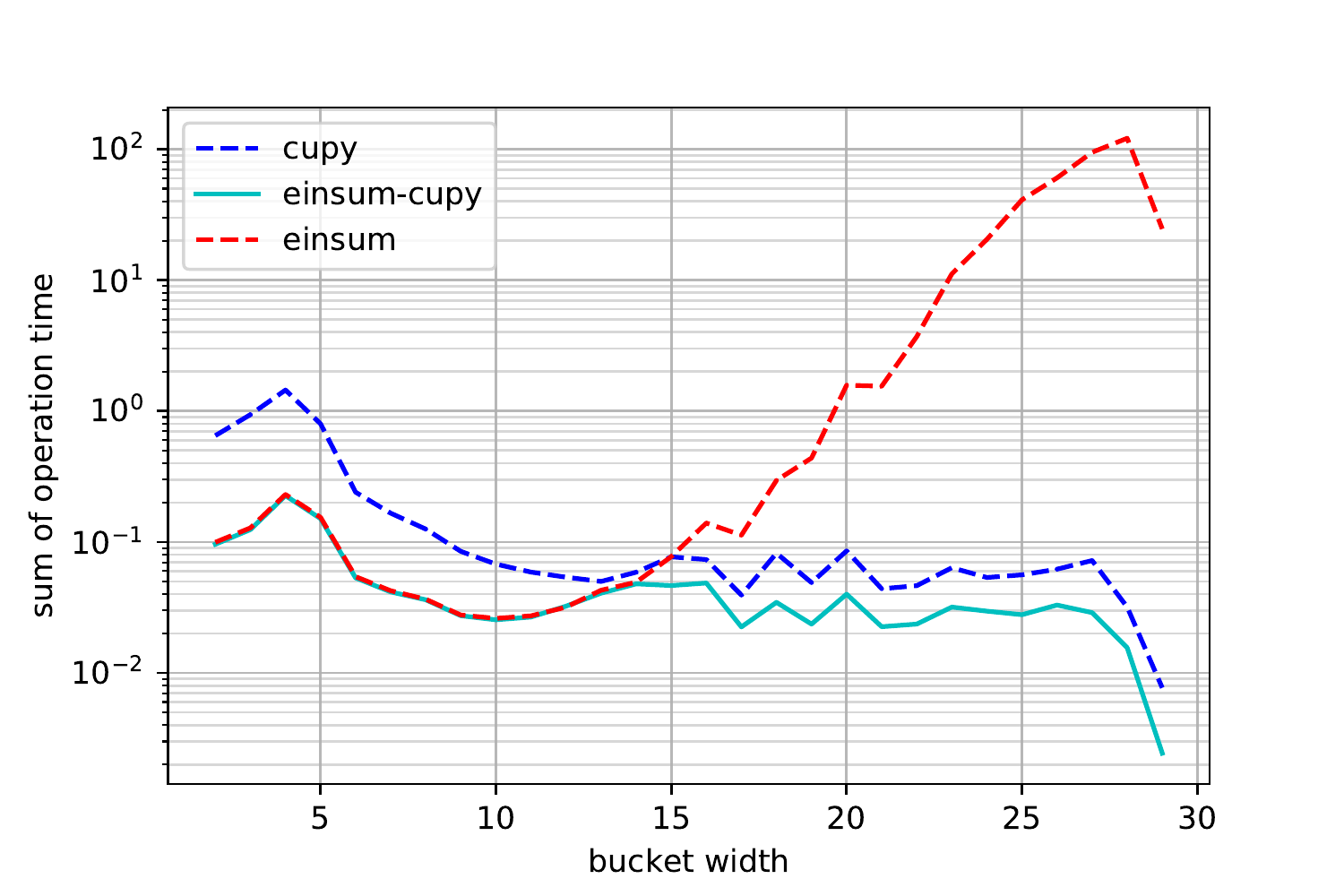}
    \caption{
    Breakdown of sum contraction time by bucket width for merged backends.
    CPU backends are better for buckets of width~$\leq15$, and GPU backends
    are better for larger buckets. The hybrid backend's GPU backend spends outperforms the regular GPU backend for buckets of width~$\geq15$.
    }
    \label{fig:meantime_bucket_MM}
\end{figure}

In Fig. \ref{fig:meantime_bucket_MM}, CPU performance is better than GPU performance when the bucket width is approximately less than 15. After 15, GPU performance scales with width much better than that of CPU performance, providing a significant speed boost over the CPU in the end. GPU performance of the hybrid backend is better than that of pure GPU backend for buckets of width ~$\geq15$. This speedup of the hybrid backend is likely caused by less garbage handling for the GPU since most buckets aren't stored on GPU memory.

\subsection{Synthetic Benchmarks}
\label{sec:results_synthetic}
We also benchmark the time required for the basic operations: matrix multiplication, tensor network contraction with fixed contraction indices, and tensor network contraction with random indices, as well as circuit contractions.

The summary of the results is shown in Table \ref{tab:summary}, which compares FLOPs count for similar-sized problems of different types.
Figures \ref{fig:tncontract_cupy} and \ref{fig:tncontract_numpy} show dependence of FLOPs vs problem size for different problems. We observe 80\% of theoretical peak performance on GPU for matrix multiplication. 
Switching to pairwise tensor network contraction shows similar FLOPs for GPU, while for CPU, it results in 10$\times$ FLOPs decrease. 
A significant reduction in performance comes from switching from pairwise tensor network contractions of a tensor with few dimensions of large size to tensors with many permuted dimensions and small size. This reduction in performance is about 10$\times$ for both CPU and GPU. This observation suggests that further improvement can be achieved by reformulating the tensor network operations in a smaller tensor by transposing and merging the dimensions of participating tensors.
It is partially addressed in using the merged indices approach, where the contraction dimension is increased.
The ``Bucket Contraction Merged" task shows 45\% of theoretical peak performance, which significantly improves compared to the unmerged counterpart.

The significant reduction of performance comes when we compare bucket contraction and full circuit contraction. 
It was explained in detail in Section \ref{sec:single_backends} and is caused by overhead from small buckets. It is evident from Figure \ref{fig:time_bucket} that most of the time in GPU simulation is spent on overhead from small bucket contraction. This issue is addressed by implementing the mixed backend approach.

It is also notable that the merged approach does not improve the performance for CPU backends which is probably due to an inefficient implementation of original \texttt{numpy.einsum()}.

\begin{table}[]
    \centering
        \begin{tabular}{|l|l|l|l|}
        \hline
         \textbf{Task}     & \textbf{CPU FLOPs}     & \textbf{GPU FLOPs}\\
         \hline
         Matrix Multiplication  &50.1G    & 2.38T\\
        \hline
         Tensor Network Fixed Contraction   &5.53G      & 1.36T\\
         \hline
         Tensor Network Random Contraction  &640M    & 97.5G\\
         \hline
         Bucket Contraction Unmerged   &241M   & 61.9G\\
         \hline
         Bucket Contraction Merged   &542M     & 1.14T\\
         \hline
         Lightcone Contraction Unmerged   &326M   & 4.92G\\
         \hline
         Lightcone Contraction Merged   &177M     & 3.1G\\
         \hline
         Circuit Contraction Mixed  &  \multicolumn{2}{c|}{ 30.7G}\\
         \hline
    \end{tabular}
    \vspace{.5em}
    \small
    \caption{\small
    Summary of GPU and CPU FLOPs for different tasks at around 100 million operations. Matrix Multiplication and Tensor Contraction tasks are described in Section \ref{sec:results_synthetic}.
    ``Bucket Contraction" groups record the maximum number of FLOPs for a single bucket. ``Lightcone Contraction" groups contain the FLOPs data on a single lightcone where the sum of operations is approximately 100 millions, small and large buckets combined.
     }
    \label{tab:summary}
\end{table}

\subsubsection{Matrix Multiplication}
The multiplication of square matrices of size 465 needs approximately 100 million complex  operations according to our calculation of operations value. The average operation time for the multiplication of two randomly generated \texttt{complex128} square matrices of size 465 is 0.3 ms on the GPU, which achieves 50$\times$ speedup compared with the operation time of 16 ms on the CPU; NumPy produces 50G FLOPs on CPU, and the GPU backend CuPy reaches 2.38T FLOPs for this operation.
We observe that the CPU backend has an advantage in performing  small operations: for matrices of size $10 \times 10$, the CPU backend NumPy  spends only 5.8 µs for the multiplication, while the best GPU backend PyTorch spends 27 µs on the operation. When the matrix size is less than $2000 \times 2000$ for the GPU backends, PyTorch outperforms CuPy, and CuPy is slightly better for much larger operations. Moreover, the operation time for both CPU and GPU backends decreases slightly when the size of matrices increases from 1000 to 1024 and from 4090 to 4096.

\subsubsection{Fixed Tensor Network Contraction}
We use the fixed contraction formula ``abcd,bcdf$\xrightarrow{}$acf'' and control the size of the tensor indices from 10 to 100. Even for the smallest case when the number of operations is 100,000 with indices of size 10, the slowest GPU backend is faster than the CPU backend Numpy, which spends 0.3 ms on the contraction. For the GPU backends, we achieve 1.36T FLOPs for this fixed contraction, which is 57\% of the recorded peak performance. In accordance with the matrix multiplication results, the CuPy backend performs better than the PyTorch backend in the fixed tensor network contractions only when the number of operations is greater than 1G.
\subsubsection{Random Tensor Network Contraction}
\begin{figure}
        \centering
        \includegraphics[width=\linewidth]{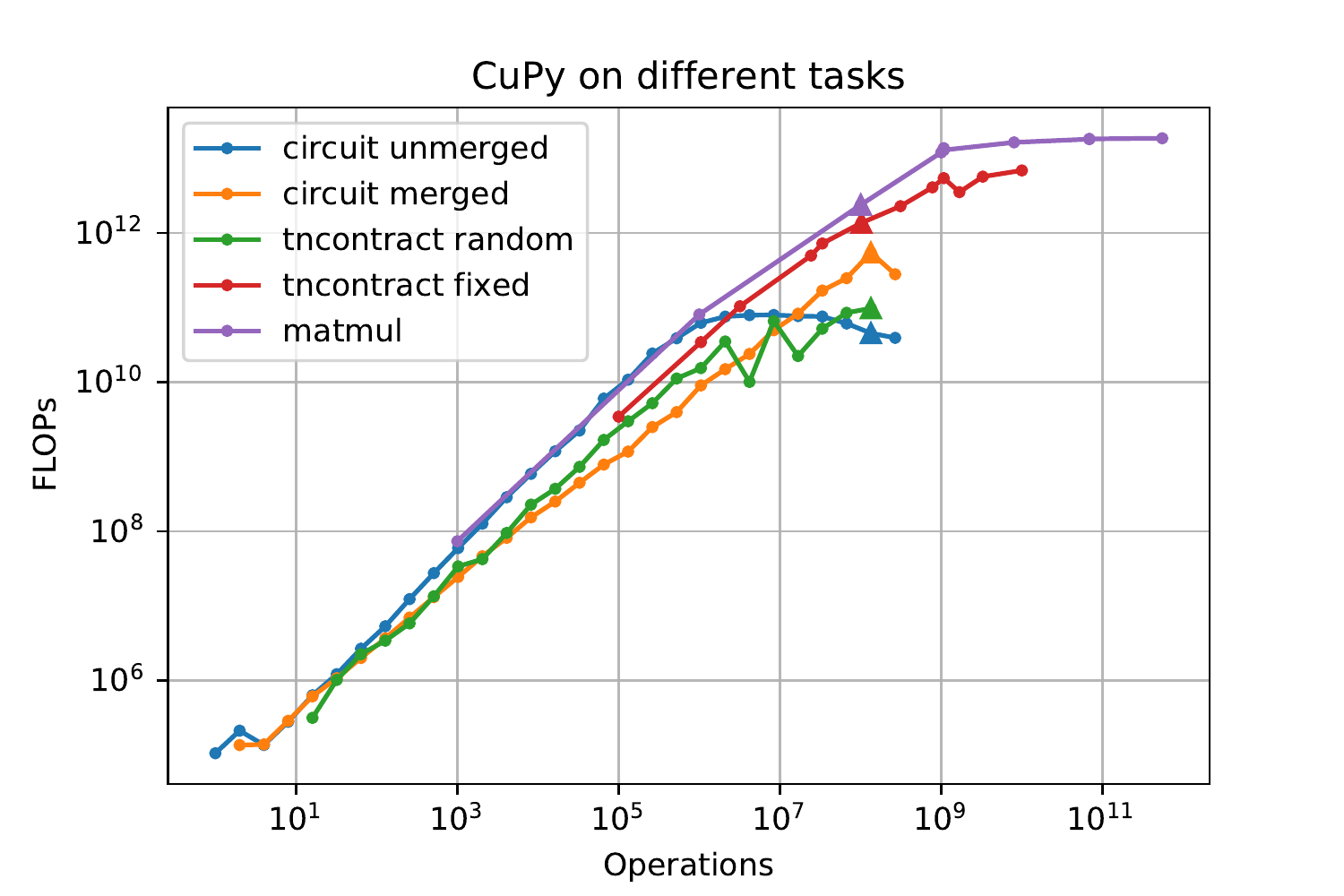}
        \caption{
        FLOPs vs. the number of operations for all tasks on the CuPy backend. ``circuit unmerged" and ``circuit merged" are results of expectation value of the full circuit simulation of QAOA MaxCut problem on a 3-regular graph of size 30 with depth $p=4$. 
        “tncontract random” tests on tensors of many indices where each index has a small size. ``tncontract fixed" uses the contraction sequence “abcd,bcdf$\xrightarrow{}$acf” for all contractions. ``matmul" performs matrix multiplication on square matrices. All groups use \texttt{complex128} tensors in the operation. We use the triangles to denote the data at $\sim100$ million operations, which is shown in Table \ref{tab:summary}.
        }
        \label{fig:tncontract_cupy}
    \end{figure}

\begin{figure}
        \centering
        \includegraphics[width=\linewidth]{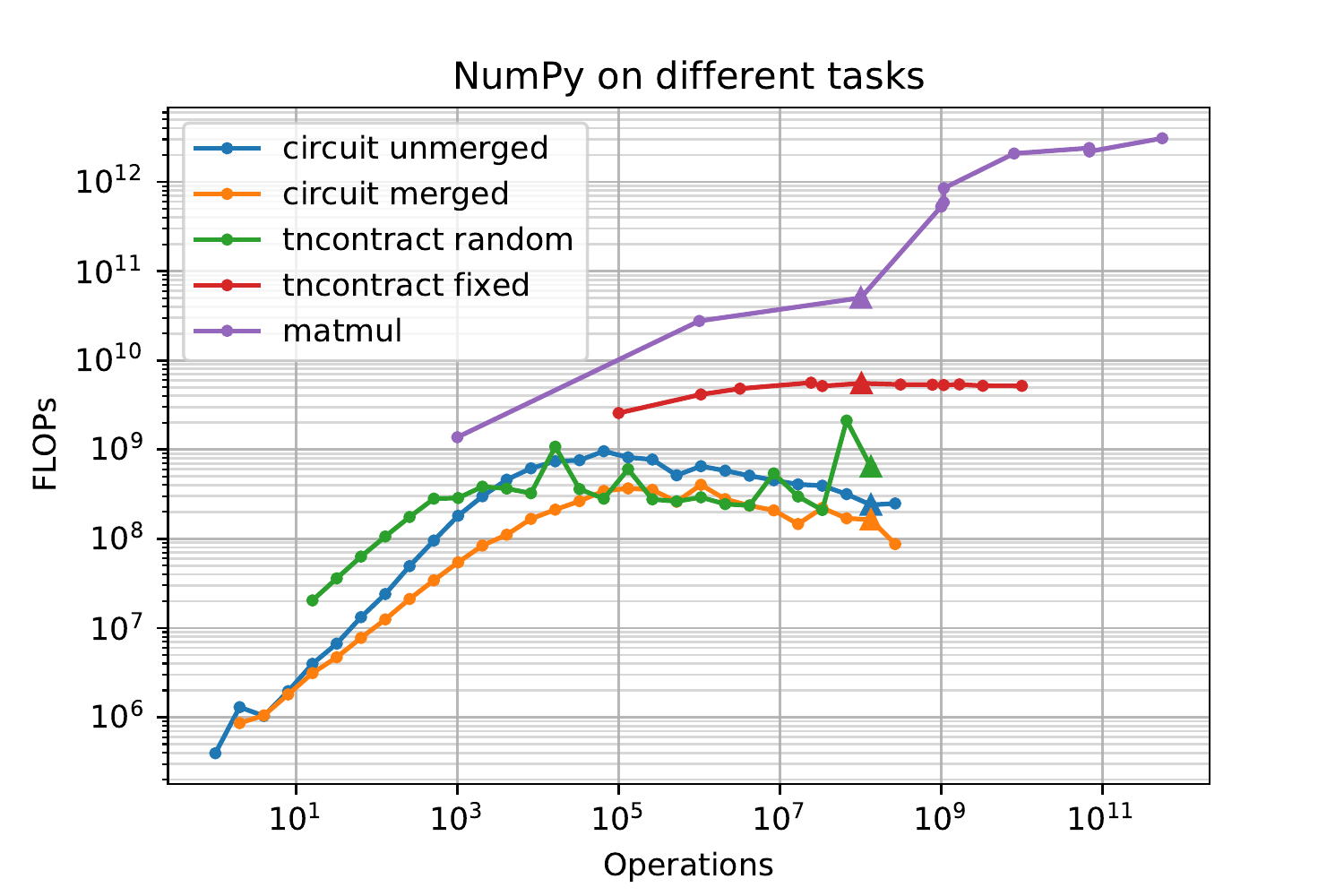}
        \caption{
        FLOPs vs. the number of operations for all tasks on NumPy backend. Same problem setting as Fig.~\ref{fig:tncontract_cupy}. ``tncontract random" outperforms ``tncontract fixed" as the ops value increases. Merged backend does not have an advantage on CPU compared to the unmerged backend.
        We use the triangles to denote the data at $\sim100$ million operations, which is shown in Table \ref{tab:summary}.
        }
        \label{fig:tncontract_numpy}
    \end{figure}

We let the number of indices be any number between 4 and 25, and we set the size of each mode to be 2. For example, we have 5 indices in total, and we randomly generate a contraction sequence ``caedb,eab$\xrightarrow{}$cde,'' so the sizes of the input tensors are $2^5$ and $2^3$, resulting in an output tensor of size $2^3$. We  reach 97.5 G FLOPs for the GPU backends and 640 M for the CPU backend only when performing this random contraction. 
As shown in Fig. \ref{fig:tncontract_cupy}, the mean FLOPs drop significantly  when we use random contraction (in green) instead of fixed contraction (in red) on the CuPy backend. On the CPU, the gap increases with the increasing number of operations according to Fig. \ref{fig:tncontract_numpy}. Therefore, contractions on tensors with small numbers of indices of large size have better performance than contractions on tensors with many indices of small size. 
The "tncontract random" group is designed to break down the circuit simulation to tensor contraction operations, so it overlaps with the results from the "bucket unmerged" group in Fig. \ref{fig:tncontract_cupy}. From the difference in performance of the random and the fixed tensor contraction group, we design the merged bucket group to improve the performance of contractions. Our goal is to make the bucket simulation curve close to the tensor contraction fixed group (the red curve).


\section{Conclusions}
This work has demonstrated that GPUs can significantly speed up quantum circuit simulations
using tensor network contractions. 
We demonstrate that GPUs are best for contracting large tensors, while CPUs are slightly better for small tensors.
Moving the computation onto GPUs can dramatically speed up the computation.
We propose to use a contraction backend that dynamically assigns the  CPU or GPU device to
tensors based on their size.
This mixed backend approach demonstrated a 176$\times$ improvement in time to solution.

We observe up to 300$\times$ speedup on GPU compared to CPU for individual large buckets.
In general, if the maximum bucketwidth of a lightcone is less than $\sim17$, the improvement from using  GPUs  is  marginal. It underlines the importance of using a mixed CPU/GPU backend for tensor contraction and using device selection for the tensor at runtime to achieve the maximum performance.  On NVIDIA DGX-2 server we found out that the threshold is $\sim15$, but it may change for other computing systems.

We also demonstrated the performance of the merged indices approach, which improves the arithmetic intensity and provides a significant FLOP improvement. Our synthetic benchmarks for various tensor contraction tasks suggest that additional improvement can be obtained by transposing and reshaping tensors in pairwise contractions.

The main conclusion of this paper is that we found that GPUs can dramatically increase the speed of tensor contractions for large tensors. The smaller tensors need to be computed on a CPU only because of overhead to move on and off data to a GPU. We show that the approach of merged indices allows to speed up large tensors contraction, but it does not solve the problem completely. Where to compute tensors leads to the problem of optimal load balancing between CPU and GPU. This potential issue will be the subject of our future work, as well as testing of the performance of the code on new NVidia DGX systems and GPU supercomputers using cuTensor and cuQuantum software packages developed by NVidia.




\section*{Acknowledgments}
Danylo Lykov and Yuri Alexeev are supported by the Defense Advanced Research Projects Agency (DARPA) grant. Yuri Alexeev and Angela Chen are also supported in a part by the Exascale Computing Project (17-SC-20-SC), a joint project of the U.S. Department of Energy’s Office of Science and National Nuclear Security Administration, responsible for delivering a capable exascale ecosystem, including software, applications, and hardware technology, to support the nation’s exascale computing imperative. Huaxuan Chen is supported in part by the U.S. Department of Energy, Office of Science, Office of Workforce Development for Teachers and Scientists (WDTS) under the Science Undergraduate Laboratory Internships Program (SULI).
This work used the resources of the Argonne Leadership Computing Facility, which is DOE Office of Science User Facility supported under Contract DE-AC02-06CH11357.

\bibliographystyle{IEEEtran}
\bibliography{qis}

\end{document}